\pdfoutput=1
\documentclass{IEEEcsmag}

\usepackage[colorlinks,urlcolor=blue,linkcolor=blue,citecolor=blue]{hyperref}

\usepackage{upmath}

\jvol{XX}
\jnum{XX}
\paper{}
\jmonth{}
\jname{IEEE Computer}
\pubyear{2023}

\let\labelindent\relax
\usepackage[inline]{enumitem}
\usepackage{cite}
\usepackage{amsmath,amssymb,amsfonts}
\usepackage{algorithmic}
\usepackage{graphicx}
\usepackage{textcomp}
\usepackage{xcolor}
\usepackage{booktabs}
\usepackage{pifont}
\usepackage{mathtools}
\usepackage{xspace}
\usepackage{caption}
\usepackage[draft]{minted}
\definecolor{bg}{rgb}{0.95,0.95,0.95}

\makeatletter
\def\PYG@reset{\let\PYG@it=\relax \let\PYG@bf=\relax%
    \let\PYG@ul=\relax \let\PYG@tc=\relax%
    \let\PYG@bc=\relax \let\PYG@ff=\relax}
\def\PYG@tok#1{\csname PYG@tok@#1\endcsname}
\def\PYG@toks#1+{\ifx\relax#1\empty\else%
    \PYG@tok{#1}\expandafter\PYG@toks\fi}
\def\PYG@do#1{\PYG@bc{\PYG@tc{\PYG@ul{%
    \PYG@it{\PYG@bf{\PYG@ff{#1}}}}}}}
\def\PYG#1#2{\PYG@reset\PYG@toks#1+\relax+\PYG@do{#2}}

\expandafter\def\csname PYG@tok@gd\endcsname{\def\PYG@tc##1{\textcolor[rgb]{0.63,0.00,0.00}{##1}}}
\expandafter\def\csname PYG@tok@gu\endcsname{\let\PYG@bf=\textbf\def\PYG@tc##1{\textcolor[rgb]{0.50,0.00,0.50}{##1}}}
\expandafter\def\csname PYG@tok@gt\endcsname{\def\PYG@tc##1{\textcolor[rgb]{0.00,0.27,0.87}{##1}}}
\expandafter\def\csname PYG@tok@gs\endcsname{\let\PYG@bf=\textbf}
\expandafter\def\csname PYG@tok@gr\endcsname{\def\PYG@tc##1{\textcolor[rgb]{1.00,0.00,0.00}{##1}}}
\expandafter\def\csname PYG@tok@cm\endcsname{\let\PYG@it=\textit\def\PYG@tc##1{\textcolor[rgb]{0.25,0.50,0.50}{##1}}}
\expandafter\def\csname PYG@tok@vg\endcsname{\def\PYG@tc##1{\textcolor[rgb]{0.10,0.09,0.49}{##1}}}
\expandafter\def\csname PYG@tok@vi\endcsname{\def\PYG@tc##1{\textcolor[rgb]{0.10,0.09,0.49}{##1}}}
\expandafter\def\csname PYG@tok@vm\endcsname{\def\PYG@tc##1{\textcolor[rgb]{0.10,0.09,0.49}{##1}}}
\expandafter\def\csname PYG@tok@mh\endcsname{\def\PYG@tc##1{\textcolor[rgb]{0.40,0.40,0.40}{##1}}}
\expandafter\def\csname PYG@tok@cs\endcsname{\let\PYG@it=\textit\def\PYG@tc##1{\textcolor[rgb]{0.25,0.50,0.50}{##1}}}
\expandafter\def\csname PYG@tok@ge\endcsname{\let\PYG@it=\textit}
\expandafter\def\csname PYG@tok@vc\endcsname{\def\PYG@tc##1{\textcolor[rgb]{0.10,0.09,0.49}{##1}}}
\expandafter\def\csname PYG@tok@il\endcsname{\def\PYG@tc##1{\textcolor[rgb]{0.40,0.40,0.40}{##1}}}
\expandafter\def\csname PYG@tok@go\endcsname{\def\PYG@tc##1{\textcolor[rgb]{0.53,0.53,0.53}{##1}}}
\expandafter\def\csname PYG@tok@cp\endcsname{\def\PYG@tc##1{\textcolor[rgb]{0.74,0.48,0.00}{##1}}}
\expandafter\def\csname PYG@tok@gi\endcsname{\def\PYG@tc##1{\textcolor[rgb]{0.00,0.63,0.00}{##1}}}
\expandafter\def\csname PYG@tok@gh\endcsname{\let\PYG@bf=\textbf\def\PYG@tc##1{\textcolor[rgb]{0.00,0.00,0.50}{##1}}}
\expandafter\def\csname PYG@tok@ni\endcsname{\let\PYG@bf=\textbf\def\PYG@tc##1{\textcolor[rgb]{0.60,0.60,0.60}{##1}}}
\expandafter\def\csname PYG@tok@nl\endcsname{\def\PYG@tc##1{\textcolor[rgb]{0.63,0.63,0.00}{##1}}}
\expandafter\def\csname PYG@tok@nn\endcsname{\let\PYG@bf=\textbf\def\PYG@tc##1{\textcolor[rgb]{0.00,0.00,1.00}{##1}}}
\expandafter\def\csname PYG@tok@no\endcsname{\def\PYG@tc##1{\textcolor[rgb]{0.53,0.00,0.00}{##1}}}
\expandafter\def\csname PYG@tok@na\endcsname{\def\PYG@tc##1{\textcolor[rgb]{0.49,0.56,0.16}{##1}}}
\expandafter\def\csname PYG@tok@nb\endcsname{\def\PYG@tc##1{\textcolor[rgb]{0.00,0.50,0.00}{##1}}}
\expandafter\def\csname PYG@tok@nc\endcsname{\let\PYG@bf=\textbf\def\PYG@tc##1{\textcolor[rgb]{0.00,0.00,1.00}{##1}}}
\expandafter\def\csname PYG@tok@nd\endcsname{\def\PYG@tc##1{\textcolor[rgb]{0.67,0.13,1.00}{##1}}}
\expandafter\def\csname PYG@tok@ne\endcsname{\let\PYG@bf=\textbf\def\PYG@tc##1{\textcolor[rgb]{0.82,0.25,0.23}{##1}}}
\expandafter\def\csname PYG@tok@nf\endcsname{\def\PYG@tc##1{\textcolor[rgb]{0.00,0.00,1.00}{##1}}}
\expandafter\def\csname PYG@tok@si\endcsname{\let\PYG@bf=\textbf\def\PYG@tc##1{\textcolor[rgb]{0.73,0.40,0.53}{##1}}}
\expandafter\def\csname PYG@tok@s2\endcsname{\def\PYG@tc##1{\textcolor[rgb]{0.73,0.13,0.13}{##1}}}
\expandafter\def\csname PYG@tok@nt\endcsname{\let\PYG@bf=\textbf\def\PYG@tc##1{\textcolor[rgb]{0.00,0.50,0.00}{##1}}}
\expandafter\def\csname PYG@tok@nv\endcsname{\def\PYG@tc##1{\textcolor[rgb]{0.10,0.09,0.49}{##1}}}
\expandafter\def\csname PYG@tok@s1\endcsname{\def\PYG@tc##1{\textcolor[rgb]{0.73,0.13,0.13}{##1}}}
\expandafter\def\csname PYG@tok@dl\endcsname{\def\PYG@tc##1{\textcolor[rgb]{0.73,0.13,0.13}{##1}}}
\expandafter\def\csname PYG@tok@ch\endcsname{\let\PYG@it=\textit\def\PYG@tc##1{\textcolor[rgb]{0.25,0.50,0.50}{##1}}}
\expandafter\def\csname PYG@tok@m\endcsname{\def\PYG@tc##1{\textcolor[rgb]{0.40,0.40,0.40}{##1}}}
\expandafter\def\csname PYG@tok@gp\endcsname{\let\PYG@bf=\textbf\def\PYG@tc##1{\textcolor[rgb]{0.00,0.00,0.50}{##1}}}
\expandafter\def\csname PYG@tok@sh\endcsname{\def\PYG@tc##1{\textcolor[rgb]{0.73,0.13,0.13}{##1}}}
\expandafter\def\csname PYG@tok@ow\endcsname{\let\PYG@bf=\textbf\def\PYG@tc##1{\textcolor[rgb]{0.67,0.13,1.00}{##1}}}
\expandafter\def\csname PYG@tok@sx\endcsname{\def\PYG@tc##1{\textcolor[rgb]{0.00,0.50,0.00}{##1}}}
\expandafter\def\csname PYG@tok@bp\endcsname{\def\PYG@tc##1{\textcolor[rgb]{0.00,0.50,0.00}{##1}}}
\expandafter\def\csname PYG@tok@c1\endcsname{\let\PYG@it=\textit\def\PYG@tc##1{\textcolor[rgb]{0.25,0.50,0.50}{##1}}}
\expandafter\def\csname PYG@tok@fm\endcsname{\def\PYG@tc##1{\textcolor[rgb]{0.00,0.00,1.00}{##1}}}
\expandafter\def\csname PYG@tok@o\endcsname{\def\PYG@tc##1{\textcolor[rgb]{0.40,0.40,0.40}{##1}}}
\expandafter\def\csname PYG@tok@kc\endcsname{\let\PYG@bf=\textbf\def\PYG@tc##1{\textcolor[rgb]{0.00,0.50,0.00}{##1}}}
\expandafter\def\csname PYG@tok@c\endcsname{\let\PYG@it=\textit\def\PYG@tc##1{\textcolor[rgb]{0.25,0.50,0.50}{##1}}}
\expandafter\def\csname PYG@tok@mf\endcsname{\def\PYG@tc##1{\textcolor[rgb]{0.40,0.40,0.40}{##1}}}
\expandafter\def\csname PYG@tok@err\endcsname{\def\PYG@bc##1{\setlength{\fboxsep}{0pt}\fcolorbox[rgb]{1.00,0.00,0.00}{1,1,1}{\strut ##1}}}
\expandafter\def\csname PYG@tok@mb\endcsname{\def\PYG@tc##1{\textcolor[rgb]{0.40,0.40,0.40}{##1}}}
\expandafter\def\csname PYG@tok@ss\endcsname{\def\PYG@tc##1{\textcolor[rgb]{0.10,0.09,0.49}{##1}}}
\expandafter\def\csname PYG@tok@sr\endcsname{\def\PYG@tc##1{\textcolor[rgb]{0.73,0.40,0.53}{##1}}}
\expandafter\def\csname PYG@tok@mo\endcsname{\def\PYG@tc##1{\textcolor[rgb]{0.40,0.40,0.40}{##1}}}
\expandafter\def\csname PYG@tok@kd\endcsname{\let\PYG@bf=\textbf\def\PYG@tc##1{\textcolor[rgb]{0.00,0.50,0.00}{##1}}}
\expandafter\def\csname PYG@tok@mi\endcsname{\def\PYG@tc##1{\textcolor[rgb]{0.40,0.40,0.40}{##1}}}
\expandafter\def\csname PYG@tok@kn\endcsname{\let\PYG@bf=\textbf\def\PYG@tc##1{\textcolor[rgb]{0.00,0.50,0.00}{##1}}}
\expandafter\def\csname PYG@tok@cpf\endcsname{\let\PYG@it=\textit\def\PYG@tc##1{\textcolor[rgb]{0.25,0.50,0.50}{##1}}}
\expandafter\def\csname PYG@tok@kr\endcsname{\let\PYG@bf=\textbf\def\PYG@tc##1{\textcolor[rgb]{0.00,0.50,0.00}{##1}}}
\expandafter\def\csname PYG@tok@s\endcsname{\def\PYG@tc##1{\textcolor[rgb]{0.73,0.13,0.13}{##1}}}
\expandafter\def\csname PYG@tok@kp\endcsname{\def\PYG@tc##1{\textcolor[rgb]{0.00,0.50,0.00}{##1}}}
\expandafter\def\csname PYG@tok@w\endcsname{\def\PYG@tc##1{\textcolor[rgb]{0.73,0.73,0.73}{##1}}}
\expandafter\def\csname PYG@tok@kt\endcsname{\def\PYG@tc##1{\textcolor[rgb]{0.69,0.00,0.25}{##1}}}
\expandafter\def\csname PYG@tok@sc\endcsname{\def\PYG@tc##1{\textcolor[rgb]{0.73,0.13,0.13}{##1}}}
\expandafter\def\csname PYG@tok@sb\endcsname{\def\PYG@tc##1{\textcolor[rgb]{0.73,0.13,0.13}{##1}}}
\expandafter\def\csname PYG@tok@sa\endcsname{\def\PYG@tc##1{\textcolor[rgb]{0.73,0.13,0.13}{##1}}}
\expandafter\def\csname PYG@tok@k\endcsname{\let\PYG@bf=\textbf\def\PYG@tc##1{\textcolor[rgb]{0.00,0.50,0.00}{##1}}}
\expandafter\def\csname PYG@tok@se\endcsname{\let\PYG@bf=\textbf\def\PYG@tc##1{\textcolor[rgb]{0.73,0.40,0.13}{##1}}}
\expandafter\def\csname PYG@tok@sd\endcsname{\let\PYG@it=\textit\def\PYG@tc##1{\textcolor[rgb]{0.73,0.13,0.13}{##1}}}

\def\PYGZus{\char`\_}
\def\PYGZob{\char`\{}
\def\PYGZcb{\char`\}}

\def\PYGZlt{\char`\<}
\def\PYGZgt{\char`\>}
\def\PYGZsh{\char`\#}

\def\PYGZdq{\char`\"}

\makeatother

\makeatletter
\def\PYGdefault@reset{\let\PYGdefault@it=\relax \let\PYGdefault@bf=\relax%
    \let\PYGdefault@ul=\relax \let\PYGdefault@tc=\relax%
    \let\PYGdefault@bc=\relax \let\PYGdefault@ff=\relax}
\def\PYGdefault@tok#1{\csname PYGdefault@tok@#1\endcsname}
\def\PYGdefault@toks#1+{\ifx\relax#1\empty\else%
    \PYGdefault@tok{#1}\expandafter\PYGdefault@toks\fi}
\def\PYGdefault@do#1{\PYGdefault@bc{\PYGdefault@tc{\PYGdefault@ul{%
    \PYGdefault@it{\PYGdefault@bf{\PYGdefault@ff{#1}}}}}}}
\def\PYGdefault#1#2{\PYGdefault@reset\PYGdefault@toks#1+\relax+\PYGdefault@do{#2}}

\expandafter\def\csname PYGdefault@tok@gd\endcsname{\def\PYGdefault@tc##1{\textcolor[rgb]{0.63,0.00,0.00}{##1}}}
\expandafter\def\csname PYGdefault@tok@gu\endcsname{\let\PYGdefault@bf=\textbf\def\PYGdefault@tc##1{\textcolor[rgb]{0.50,0.00,0.50}{##1}}}
\expandafter\def\csname PYGdefault@tok@gt\endcsname{\def\PYGdefault@tc##1{\textcolor[rgb]{0.00,0.27,0.87}{##1}}}
\expandafter\def\csname PYGdefault@tok@gs\endcsname{\let\PYGdefault@bf=\textbf}
\expandafter\def\csname PYGdefault@tok@gr\endcsname{\def\PYGdefault@tc##1{\textcolor[rgb]{1.00,0.00,0.00}{##1}}}
\expandafter\def\csname PYGdefault@tok@cm\endcsname{\let\PYGdefault@it=\textit\def\PYGdefault@tc##1{\textcolor[rgb]{0.25,0.50,0.50}{##1}}}
\expandafter\def\csname PYGdefault@tok@vg\endcsname{\def\PYGdefault@tc##1{\textcolor[rgb]{0.10,0.09,0.49}{##1}}}
\expandafter\def\csname PYGdefault@tok@vi\endcsname{\def\PYGdefault@tc##1{\textcolor[rgb]{0.10,0.09,0.49}{##1}}}
\expandafter\def\csname PYGdefault@tok@vm\endcsname{\def\PYGdefault@tc##1{\textcolor[rgb]{0.10,0.09,0.49}{##1}}}
\expandafter\def\csname PYGdefault@tok@mh\endcsname{\def\PYGdefault@tc##1{\textcolor[rgb]{0.40,0.40,0.40}{##1}}}
\expandafter\def\csname PYGdefault@tok@cs\endcsname{\let\PYGdefault@it=\textit\def\PYGdefault@tc##1{\textcolor[rgb]{0.25,0.50,0.50}{##1}}}
\expandafter\def\csname PYGdefault@tok@ge\endcsname{\let\PYGdefault@it=\textit}
\expandafter\def\csname PYGdefault@tok@vc\endcsname{\def\PYGdefault@tc##1{\textcolor[rgb]{0.10,0.09,0.49}{##1}}}
\expandafter\def\csname PYGdefault@tok@il\endcsname{\def\PYGdefault@tc##1{\textcolor[rgb]{0.40,0.40,0.40}{##1}}}
\expandafter\def\csname PYGdefault@tok@go\endcsname{\def\PYGdefault@tc##1{\textcolor[rgb]{0.53,0.53,0.53}{##1}}}
\expandafter\def\csname PYGdefault@tok@cp\endcsname{\def\PYGdefault@tc##1{\textcolor[rgb]{0.74,0.48,0.00}{##1}}}
\expandafter\def\csname PYGdefault@tok@gi\endcsname{\def\PYGdefault@tc##1{\textcolor[rgb]{0.00,0.63,0.00}{##1}}}
\expandafter\def\csname PYGdefault@tok@gh\endcsname{\let\PYGdefault@bf=\textbf\def\PYGdefault@tc##1{\textcolor[rgb]{0.00,0.00,0.50}{##1}}}
\expandafter\def\csname PYGdefault@tok@ni\endcsname{\let\PYGdefault@bf=\textbf\def\PYGdefault@tc##1{\textcolor[rgb]{0.60,0.60,0.60}{##1}}}
\expandafter\def\csname PYGdefault@tok@nl\endcsname{\def\PYGdefault@tc##1{\textcolor[rgb]{0.63,0.63,0.00}{##1}}}
\expandafter\def\csname PYGdefault@tok@nn\endcsname{\let\PYGdefault@bf=\textbf\def\PYGdefault@tc##1{\textcolor[rgb]{0.00,0.00,1.00}{##1}}}
\expandafter\def\csname PYGdefault@tok@no\endcsname{\def\PYGdefault@tc##1{\textcolor[rgb]{0.53,0.00,0.00}{##1}}}
\expandafter\def\csname PYGdefault@tok@na\endcsname{\def\PYGdefault@tc##1{\textcolor[rgb]{0.49,0.56,0.16}{##1}}}
\expandafter\def\csname PYGdefault@tok@nb\endcsname{\def\PYGdefault@tc##1{\textcolor[rgb]{0.00,0.50,0.00}{##1}}}
\expandafter\def\csname PYGdefault@tok@nc\endcsname{\let\PYGdefault@bf=\textbf\def\PYGdefault@tc##1{\textcolor[rgb]{0.00,0.00,1.00}{##1}}}
\expandafter\def\csname PYGdefault@tok@nd\endcsname{\def\PYGdefault@tc##1{\textcolor[rgb]{0.67,0.13,1.00}{##1}}}
\expandafter\def\csname PYGdefault@tok@ne\endcsname{\let\PYGdefault@bf=\textbf\def\PYGdefault@tc##1{\textcolor[rgb]{0.82,0.25,0.23}{##1}}}
\expandafter\def\csname PYGdefault@tok@nf\endcsname{\def\PYGdefault@tc##1{\textcolor[rgb]{0.00,0.00,1.00}{##1}}}
\expandafter\def\csname PYGdefault@tok@si\endcsname{\let\PYGdefault@bf=\textbf\def\PYGdefault@tc##1{\textcolor[rgb]{0.73,0.40,0.53}{##1}}}
\expandafter\def\csname PYGdefault@tok@s2\endcsname{\def\PYGdefault@tc##1{\textcolor[rgb]{0.73,0.13,0.13}{##1}}}
\expandafter\def\csname PYGdefault@tok@nt\endcsname{\let\PYGdefault@bf=\textbf\def\PYGdefault@tc##1{\textcolor[rgb]{0.00,0.50,0.00}{##1}}}
\expandafter\def\csname PYGdefault@tok@nv\endcsname{\def\PYGdefault@tc##1{\textcolor[rgb]{0.10,0.09,0.49}{##1}}}
\expandafter\def\csname PYGdefault@tok@s1\endcsname{\def\PYGdefault@tc##1{\textcolor[rgb]{0.73,0.13,0.13}{##1}}}
\expandafter\def\csname PYGdefault@tok@dl\endcsname{\def\PYGdefault@tc##1{\textcolor[rgb]{0.73,0.13,0.13}{##1}}}
\expandafter\def\csname PYGdefault@tok@ch\endcsname{\let\PYGdefault@it=\textit\def\PYGdefault@tc##1{\textcolor[rgb]{0.25,0.50,0.50}{##1}}}
\expandafter\def\csname PYGdefault@tok@m\endcsname{\def\PYGdefault@tc##1{\textcolor[rgb]{0.40,0.40,0.40}{##1}}}
\expandafter\def\csname PYGdefault@tok@gp\endcsname{\let\PYGdefault@bf=\textbf\def\PYGdefault@tc##1{\textcolor[rgb]{0.00,0.00,0.50}{##1}}}
\expandafter\def\csname PYGdefault@tok@sh\endcsname{\def\PYGdefault@tc##1{\textcolor[rgb]{0.73,0.13,0.13}{##1}}}
\expandafter\def\csname PYGdefault@tok@ow\endcsname{\let\PYGdefault@bf=\textbf\def\PYGdefault@tc##1{\textcolor[rgb]{0.67,0.13,1.00}{##1}}}
\expandafter\def\csname PYGdefault@tok@sx\endcsname{\def\PYGdefault@tc##1{\textcolor[rgb]{0.00,0.50,0.00}{##1}}}
\expandafter\def\csname PYGdefault@tok@bp\endcsname{\def\PYGdefault@tc##1{\textcolor[rgb]{0.00,0.50,0.00}{##1}}}
\expandafter\def\csname PYGdefault@tok@c1\endcsname{\let\PYGdefault@it=\textit\def\PYGdefault@tc##1{\textcolor[rgb]{0.25,0.50,0.50}{##1}}}
\expandafter\def\csname PYGdefault@tok@fm\endcsname{\def\PYGdefault@tc##1{\textcolor[rgb]{0.00,0.00,1.00}{##1}}}
\expandafter\def\csname PYGdefault@tok@o\endcsname{\def\PYGdefault@tc##1{\textcolor[rgb]{0.40,0.40,0.40}{##1}}}
\expandafter\def\csname PYGdefault@tok@kc\endcsname{\let\PYGdefault@bf=\textbf\def\PYGdefault@tc##1{\textcolor[rgb]{0.00,0.50,0.00}{##1}}}
\expandafter\def\csname PYGdefault@tok@c\endcsname{\let\PYGdefault@it=\textit\def\PYGdefault@tc##1{\textcolor[rgb]{0.25,0.50,0.50}{##1}}}
\expandafter\def\csname PYGdefault@tok@mf\endcsname{\def\PYGdefault@tc##1{\textcolor[rgb]{0.40,0.40,0.40}{##1}}}
\expandafter\def\csname PYGdefault@tok@err\endcsname{\def\PYGdefault@bc##1{\setlength{\fboxsep}{0pt}\fcolorbox[rgb]{1.00,0.00,0.00}{1,1,1}{\strut ##1}}}
\expandafter\def\csname PYGdefault@tok@mb\endcsname{\def\PYGdefault@tc##1{\textcolor[rgb]{0.40,0.40,0.40}{##1}}}
\expandafter\def\csname PYGdefault@tok@ss\endcsname{\def\PYGdefault@tc##1{\textcolor[rgb]{0.10,0.09,0.49}{##1}}}
\expandafter\def\csname PYGdefault@tok@sr\endcsname{\def\PYGdefault@tc##1{\textcolor[rgb]{0.73,0.40,0.53}{##1}}}
\expandafter\def\csname PYGdefault@tok@mo\endcsname{\def\PYGdefault@tc##1{\textcolor[rgb]{0.40,0.40,0.40}{##1}}}
\expandafter\def\csname PYGdefault@tok@kd\endcsname{\let\PYGdefault@bf=\textbf\def\PYGdefault@tc##1{\textcolor[rgb]{0.00,0.50,0.00}{##1}}}
\expandafter\def\csname PYGdefault@tok@mi\endcsname{\def\PYGdefault@tc##1{\textcolor[rgb]{0.40,0.40,0.40}{##1}}}
\expandafter\def\csname PYGdefault@tok@kn\endcsname{\let\PYGdefault@bf=\textbf\def\PYGdefault@tc##1{\textcolor[rgb]{0.00,0.50,0.00}{##1}}}
\expandafter\def\csname PYGdefault@tok@cpf\endcsname{\let\PYGdefault@it=\textit\def\PYGdefault@tc##1{\textcolor[rgb]{0.25,0.50,0.50}{##1}}}
\expandafter\def\csname PYGdefault@tok@kr\endcsname{\let\PYGdefault@bf=\textbf\def\PYGdefault@tc##1{\textcolor[rgb]{0.00,0.50,0.00}{##1}}}
\expandafter\def\csname PYGdefault@tok@s\endcsname{\def\PYGdefault@tc##1{\textcolor[rgb]{0.73,0.13,0.13}{##1}}}
\expandafter\def\csname PYGdefault@tok@kp\endcsname{\def\PYGdefault@tc##1{\textcolor[rgb]{0.00,0.50,0.00}{##1}}}
\expandafter\def\csname PYGdefault@tok@w\endcsname{\def\PYGdefault@tc##1{\textcolor[rgb]{0.73,0.73,0.73}{##1}}}
\expandafter\def\csname PYGdefault@tok@kt\endcsname{\def\PYGdefault@tc##1{\textcolor[rgb]{0.69,0.00,0.25}{##1}}}
\expandafter\def\csname PYGdefault@tok@sc\endcsname{\def\PYGdefault@tc##1{\textcolor[rgb]{0.73,0.13,0.13}{##1}}}
\expandafter\def\csname PYGdefault@tok@sb\endcsname{\def\PYGdefault@tc##1{\textcolor[rgb]{0.73,0.13,0.13}{##1}}}
\expandafter\def\csname PYGdefault@tok@sa\endcsname{\def\PYGdefault@tc##1{\textcolor[rgb]{0.73,0.13,0.13}{##1}}}
\expandafter\def\csname PYGdefault@tok@k\endcsname{\let\PYGdefault@bf=\textbf\def\PYGdefault@tc##1{\textcolor[rgb]{0.00,0.50,0.00}{##1}}}
\expandafter\def\csname PYGdefault@tok@se\endcsname{\let\PYGdefault@bf=\textbf\def\PYGdefault@tc##1{\textcolor[rgb]{0.73,0.40,0.13}{##1}}}
\expandafter\def\csname PYGdefault@tok@sd\endcsname{\let\PYGdefault@it=\textit\def\PYGdefault@tc##1{\textcolor[rgb]{0.73,0.13,0.13}{##1}}}

\makeatother

\usepackage{tikz}
\usetikzlibrary{shapes}
\newcommand*\circled[1]{\tikz[baseline=(char.base)]{
    \node[shape=circle,draw,inner sep=1pt] (char) {\textbf{#1}};}}

\newcommand{\takeawaybox}[4]{
  \begin{figure}[h]
    \centering
    \begin{tikzpicture}
      \node[anchor=text,text width=\columnwidth-1.2cm, draw, rounded corners, line width=1pt, fill=#3, inner sep=3mm] (big) {\\#4};
      \node[draw, rounded corners, line width=.5pt, fill=#2, anchor=west, xshift=5mm] (small) at (big.north west) {\textbf{#1}};
    \end{tikzpicture}
  \end{figure}
}

\usepackage{soul}

\usepackage{marginnote}
\usepackage{etoolbox}
\robustify{\marginnote}

\newcommand{\hlc}[1]{\colorlet{clr}{cyan!20}\sethlcolor{clr}\hl{#1}}
\robustify{\hlc}
\newcommand{\hlcy}[1]{\colorlet{clr}{orange!20}\sethlcolor{clr}\hl{#1}}
\robustify{\hlc}

\newif\ifrev
\revfalse
\newcommand{\rev}[1]{\ifrev\hlc{#1}\else#1\fi}
\newcommand{\highlight}[1]{\ifrev\hlcy{#1}\else#1\fi}
\newcommand{\mnote}[1]{\ifrev\normalmarginpar\marginnote{\textcolor{blue}{\textbf{#1}}}\fi}
\newcommand{\mnoter}[1]{\ifrev\reversemarginpar\marginnote{\textcolor{blue}{\textbf{#1}}}\fi}

\newcommand{\highlightmnoter}[1]{\ifrev\reversemarginpar\marginnote{\textcolor{orange}{\textbf{#1}}}\fi}

\soulregister{\cite}{7}
\soulregister{\ref}{7}
\soulregister{\ie}{7}
\soulregister{\eg}{7}
\soulregister{\gensync}{7}

\newcommand{\ie}{\emph{i.e.,}\xspace}
\newcommand{\eg}{\emph{e.g.,}\xspace}
\newcommand{\ea}{\emph{at al.}}

\newcommand{\gensync}{\emph{GenSync}\xspace}

\newcommand{\fref}[1]{Fig.~\ref{#1}}

\newcommand{\algo}[1]{\noindent\underline{\textbf{#1}}}

\newcommand{\concept}[1]{\noindent\underline{\textit{\textbf{#1}}}}

\begin{document}

\sptitle{Department: Head}
\editor{Editor: Name, xxxx@email}

\title{\mnoter{2.a}\rev{Enabling Cost-Benefit Analysis of Data Sync Protocols}}

\author{Novak Boškov, Ari Trachtenberg and David Starobinski}
\affil{Boston University}

\markboth{Department Head}{Paper title}

\begin{abstract}
  The problem of data synchronization arises in networked applications
that require some measure of consistency. Indeed data synchronization
approaches have demonstrated a significant potential for improving
performance in various applications ranging from distributed ledgers
to fog-enabled storage offloading for IoT. Although several
\mnoter{2.a}\rev{protocols for data sets synchronization} have been
proposed over the years, there is currently no widespread utility
implementing them, unlike the popular Rsync utility available for file
synchronization. To that end, we describe a new middleware called
\gensync{} that abstracts the subtleties \mnoter{2.a}\rev{of the
  state-of-the-art data synchronization protocols, allows users to
  choose protocols based on a comparative evaluation under realistic
  system conditions, and seamlessly integrate protocols in existing
  applications through a public API.} We showcase \gensync{} through a
case study, in which we integrate it into one of the world's largest
wireless emulators and compare the performance of its included
protocols.

\end{abstract}

\maketitle

\chapterinitial{Introduction}
Replication is a common thread among disparate distributed systems, typically
arising when there is a need for \textit{fault tolerance} or \textit{availability}.
Multiple replicas enable an administrator to reroute requests from
failed to healthy replicas, while a repair is completed.  Repaired
replicas can then be returned to a consistent state from existing
healthy replicas.

Replication can also be used as a tool to improve the
\textit{performance} of distributed systems with many
concurrent accesses. For instance, globally distributed databases and
content delivery networks use replication to speed up accesses from
geographically dispersed locations.

Beyond availability, fault tolerance and performance benefits,
replicated systems can also facilitate \textit{decentralization}. For
example, in distributed ledger technologies (\eg blockchains), each
participant holds a replica of an entire data set (\ie transactions),
which allows any participant to independently verify the state of
system. As long as enough participants maintain a substantial level of
consistency among their replicas, no one needs to trust a central
authority.

\mnoter{1.a, 1.c, 2.a}\rev{At the core of replication is \emph{data
    synchronization (sync)} or \emph{reconciliation}, the process that
  brings multiple replicas into a consistent state.}
Assuming that there are two parties aiming at
syncing their data sets, a naive solution would simply exchange data
sets. Having both sets, the parties can easily identify the differing
elements and include them in their own replicas. However, this is
prohibitively expensive, because communicating huge data
sets can be slow for bandwidth-constrained networks or
resource constrained compute platforms (\eg augmented reality and
wearables
). Worse yet, huge sets may differ
in only a few elements, making most of the communication redundant.
The waste is trebled when multiple participants exchange their sets in
a peer-to-peer fashion, as is the case for distributed ledgers.

\subsection*{\mnote{1.a, 1.c, 2.a}\rev{Better Ways to Synchronize Data}}

Many protocols have been developed over time to address the weaknesses
of the naive sync approach. Perhaps the most popular is
Tridgell and Mackerras's Rsync~\cite{tridgell}, which is
included in most Linux distributions and used as a default
network-enabled synchronization utility.
Originally designed for files (or bit arrays), Rsync operates by
dividing the files into chunks, exchanging the hashes of chunks, and
generating \emph{instructions} on how to make two files equal. These
instructions are transferred over the network and applied to synchronize the files. 


Rsync does not readily generalize from files to data sets, however,
there have been other approaches proposed for that purpose. One such
approach is somewhat similar to Rsync and uses a specialized hash
known as \emph{Bloom filter} to compactly represent the data set.
Rather than exchanging hashes of chunks like Rsync, this approach
exchanges Bloom filters. 
The main drawback of this approach, however, is
non-optimal usage of bandwidth --- it exchanges some traffic even for
the set elements evident on both sides of the communication
channel. A \emph{communication-efficient} protocol, on the other hand, should
only exchange the information related to differences between the
sets~\cite{minsky2003set}.



\subsection*{\mnote{1.a, 1.c, 2.a}\rev{Choosing the Right Sync Protocol}}

Several communication-efficient protocols have been proposed
in the literature. Some rely on coding theory, while other make use of
probabilistic data structures. These subtle but significant
differences in design make it important to be able to compare them under practical
conditions. Yet prior to our work~\cite{gensync}, there were no publicly
available general tools that afforded such a systematic comparison.  Our work
has shown that a protocol may dominate under certain network conditions,
but grossly underperform when the network conditions change. Worse yet, the best
protocol choice is a function not only of network conditions, but also
of the immediate compute capacities of the nodes.


\subsection*{\rev{Contributions}}

There are two significant obstacles for developers in choosing the
right protocol for their
applications: \begin{enumerate*}[label=(\arabic*)]
\item the lack of a utility for comparative analysis of the sync
  protocols under practical system conditions, and
\item the complexity of integrating sync protocols into existing
  implementations.
\end{enumerate*}
\mnoter{1.a, 1.c, 2.a}\rev{Therefore, the main objective of this work is to overcome these
  obstacles and enable the integration of the state-of-the art sync
  protocols in future applications,} such as
6G-enabled Enhanced Reality (ER), Internet-of-Things (IoT) and
Internet-of-Vehicles (IoV), where sync protocol customization is
needed at the application level. \rev{We summarize our contributions
  as follows}:




\begin{itemize}
\item We describe our middleware \gensync{}, \rev{to the best of our
    knowledge, the first utility} that enables a systematic
  cost-benefit analysis of utilizing different sync
  protocols\mnoter{1.f, 2.b, 3.b}.
\item We demonstrate the use of this middleware in an independent,
  large-scale wireless emulator, \emph{Colosseum}.
\item We show that the choice of sync protocols can significantly
  impair or improve performance.
\end{itemize}

\section*{\mnote{3.c}\rev{Applications of Data Sync}}\label{sec:in_the_wild}
Data sync serves as a building block for a diverse collection of
applications and computing paradigms, though it is sometimes embedded
deeply within the architecture of these
applications. \rev{Next, we describe example applications
  that can benefit from \gensync{}, focusing on distributed ledgers,
  cloud storage services, and IoT storage offloading.}\mnoter{3.c}


\subsection*{Distributed Ledgers}
Distributed ledgers record the decentralized transactions
of massive amounts of participants. The most popular among
distributed ledgers, blockchains, implement their transaction
inventory as a list of transaction blocks that are logically
chained together with cryptographic hashes.


Recent advances in blockchain and distribute ledger technologies have
opened a range of new possibilities for tackling long-standing
problems in related areas such as IoT access
management~\cite{IoTAccess}, security of federated learning in fog
computing~\cite{blockhainfedsurvey}, accident forensics in vehicular
networks (VANET) of self-driving cars~\cite{block4forensic}, or
information poisoning prevention in mission-critical unmanned aerial
vehicle networks (UAANET)~\cite{icnblockchain}. 

\mnote{2.e, 3.c, 3.e,
  3.f}\rev{Applying blockchain technology over disparate
  layers of Distributed Computing Continuum Systems
  (DCCS)~\cite{dccs}, which are characterized by heterogeneous compute
  and network resources, pose new challenges to performance and
  reliability. To cope with these challenges,
  several improvements to the blockchain's networking layer have
  recently been proposed}, based on set reconciliation protocols,
including MempoolSync~\cite{churnAnas2021},
Graphene~\cite{ozisik2019graphene}, and
Erlay~\cite{naumenko2019erlay}. The performance of these protocols has
been shown to vary significantly depending on the network conditions
and compute capabilities of nodes~\cite{gensync}, complicating the
analysis and choice of optimum reconciliation protocol and parameters.

\subsection*{Cloud Storage Services}
Cloud storage services such as Apple iCloud, Dropbox, Google Cloud,
and Microsoft OneDrive have also become commonplace for modern
internet users and are known to generate tremendous amounts of data
traffic for the cloud providers
~\cite{towardsnetleveff}. To reduce the amount of data transfer, these
applications employ protocols that are commonly referred to as
\emph{delta synchronization}
. The main objective of
these protocols is to determine and transmit only those portions of
data that have been updated locally. In that vein, the research
community has defined a metric called TUE (Traffic Usage Efficiency)
that is the ratio between total sync data traffic and the update
size~\cite{towardsnetleveff}. The state-of-the art delta
synchronization protocols currently in use 
rely on improved versions of Rsync. 

\subsection*{IoT Storage Offloading}
One of the significant problems in IoT applications is storage
offloading. A standard IoT setup deals with a relatively large
number of IoT devices that each produce a substantial amount of data
but lack storage and compute capacity to maintain it.
Traditionally, this problem has been addressed by moving data to
the cloud for processing. However, there are two significant drawbacks
of such approaches:
\begin{enumerate*}[label=(\arabic*)]
\item the data is physically transferred to another entity, which
  raises various security concerns, and
\item data access latency may be unacceptable for real-time
  applications as each data access goes through a wide-area network.
\end{enumerate*}
To tackle these drawbacks, Wang~\ea~\cite{wang2018fog} have come up
with a fog-based architecture, based on Rsync, that allows for data
storage within the boundaries of the entity that operates the IoT
devices. The proposed architecture has three layers. The lowest layer
consists of IoT devices that synchronize their data with the fog layer
above them. The fog layer accumulates this data and synchronizes it with
the cloud in batches.


\subsection*{Other Applications}

There are many other distributed systems where data sync is used as a
building block. For instance, researchers proposed a dissemination
protocol for wireless sensor networks that use a variant of Bloom
filter-based data sync to reduce energy consumption and propagation
delay~\cite{chen2009bdp}. Similarly, a physiological value-based key
agreement scheme for body area networks called E2PKA~\cite{Choi2017}
uses data sync to reduce memory footprint and energy consumption.
Data sync protocols have also been proposed as a solution to network
partitioning in information-centric networking (ICN)~\cite{chronosync}
as well as re-establishing consistency among replicas of distributed
databases~\cite{chen2014robust}. \mnoter{3.c, 3.e}\rev{As Distributed
  Computing Continuum Systems~\cite{dccs, dccs_rel} and applications
  that operate across the layers of this continuum emerge, we expect that
  many more applications could take advantage of an efficient data sync middleware.
}


\section*{\gensync{} Middleware}\label{sec:middleware_design}

\begin{figure}[h]
  \centering
  \includegraphics[width=\linewidth]{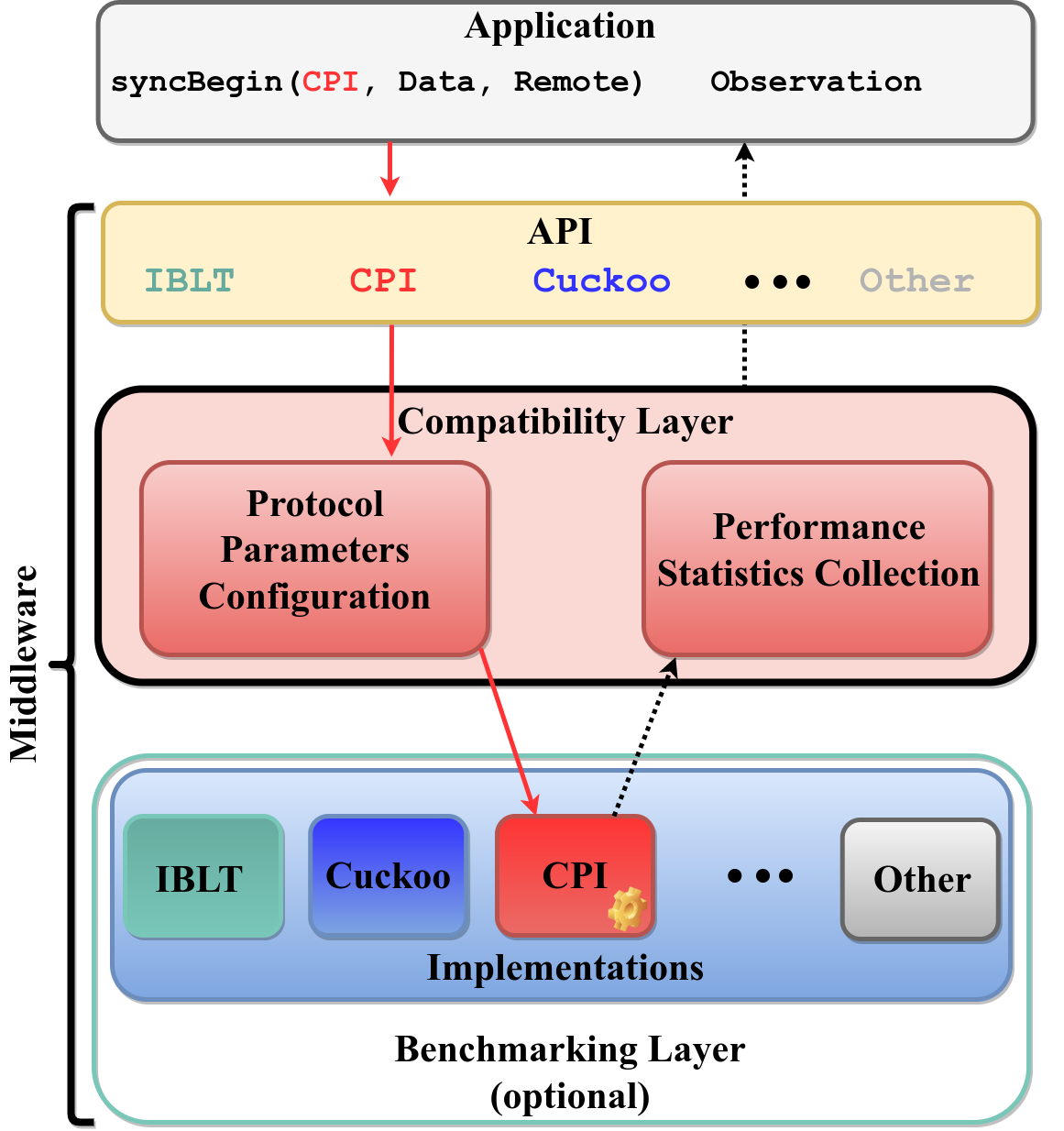}
  \caption{\mnoter{2.e, 3.d}\rev{Middleware design}.}\label{fig:middleware}
\end{figure}

We implement \gensync{} framework as a C++ middleware library that can
be used through an API, as shown in
\fref{fig:middleware}. \mnoter{3.a}\rev{The main challenges that we  tackled in designing \gensync{} were}
\begin{enumerate*}[label=(\arabic*)]
\item \rev{unifying and simplifying the parameterization of disparate
    state-of-the-art data sync protocols};
\item \rev{designing generic protocol implementations that enable
    \gensync{} utilization in various applications; and}
\item \rev{constructing a lightweight yet versatile benchmarking layer.}
\end{enumerate*}
\rev{From the \gensync{} users' standpoint}, the interaction between the
application and the middleware library is carried out through three
abstractions, named \emph{GenSync}, \emph{Observation}, and
\emph{Builder}. 



\concept{GenSync} is a generic representation of a data sync
protocol. Its interface consists of two main methods,
\emph{addElement}, which adds elements to the associated data
set, and \emph{syncBegin}, which initiates synchronization with
an external party using the desired protocol.
In a typical application, set elements can be easily mapped to a set
of unique identifiers using a \textit{hash function} and passed to
\emph{addElement}. \mnote{2.d}\rev{We design \emph{GenSync} as an
  abstract interface to allow for middleware extensions, which is
  particularly useful to researchers that design novel set
  reconciliation protocols and want to benchmark against the
  state-of-the art, and the practitioners that design
  platform-specific implementations of existing protocols}. A
custom sync implementation needs only to implement the two
\emph{GenSync} methods. By implementing \emph{addElement}, users can
control how data set is being maintained, while \emph{syncBegin} can
be used to provide the core implementation of the custom protocol.



\concept{Observation} is the summarized result of the sync and will be
generated by \emph{syncBegin}. It represents a collection of execution
statistics including the measure of success, exact protocol
parameters that have been used and monitoring information, such
as the communication cost and time expended on data transfer
and computation.

\concept{Builder} is an auxiliary abstraction that facilitates the
creation and connection of \emph{GenSyncs} to each other.  For example,
an application may want to instantiate multiple \emph{GenSync} objects
(say, one for each of several neighbors in a peer-to-peer system).  The
builder allows the application to attach each \emph{GenSync} object to
remote peers through \emph{Communicants}, which abstract out a communication
channel (\eg a TCP, UDP, or local Unix socket). The \emph{Builder} abstraction is
particularly useful in heterogeneous environments, where peers may
connect using different underlying transport or even physical-layer
protocols, or may want to use individually optimized sync protocols for
different neighbors. Code listing~\ref{listing:chain} shows how
\emph{Builder} and \emph{GenSync} can be chained together to conduct
the sync and produce an \emph{Observation}.

\begin{listing}[h]





\footnotesize
\begin{Verbatim}[commandchars=\\\{\}]
\PYG{c+cp}{\PYGZsh{}include} \PYG{c+cpf}{\PYGZlt{}GenSync.h\PYGZgt{}}

\PYG{c+c1}{// Point to a remote and pick a protocol}
\PYG{k}{auto} \PYG{n}{builder} \PYG{o}{=} \PYG{n}{GenSync}\PYG{o}{::}\PYG{n}{Builder}\PYG{p}{();}
\PYG{n}{builder}\PYG{p}{.}\PYG{n}{setProtocol}\PYG{p}{(}\PYG{n}{GenSync}\PYG{o}{::}\PYG{n}{CPI}\PYG{p}{);}
\PYG{n}{builder}\PYG{p}{.}\PYG{n}{setCommunicant}\PYG{p}{(}\PYG{n}{GenSync}\PYG{o}{::}\PYG{n}{socket}\PYG{p}{);}
\PYG{n}{builder}\PYG{p}{.}\PYG{n}{setHost}\PYG{p}{(}\PYG{l+s}{\PYGZdq{}the.peer.remote.addr\PYGZdq{}}\PYG{p}{);}

\PYG{n}{GenSync} \PYG{n}{gs} \PYG{o}{=} \PYG{n}{builder}\PYG{p}{.}\PYG{n}{build}\PYG{p}{();}

\PYG{c+c1}{// Add data}
\PYG{k}{for} \PYG{p}{(}\PYG{k}{auto} \PYG{n+nl}{data\PYGZus{}point} \PYG{p}{:} \PYG{n}{data\PYGZus{}set}\PYG{p}{)}
  \PYG{n}{gs}\PYG{p}{.}\PYG{n}{addElement}\PYG{p}{(}\PYG{n}{hash}\PYG{p}{(}\PYG{n}{data\PYGZus{}point}\PYG{p}{));}

\PYG{c+c1}{// Perform sync}
\PYG{k}{if} \PYG{p}{(} \PYG{n}{gs}\PYG{p}{.}\PYG{n}{syncBegin}\PYG{p}{()} \PYG{p}{)} \PYG{p}{\PYGZob{}}
  \PYG{c+c1}{// Get execution statistics}
  \PYG{n}{Observation} \PYG{n}{ob} \PYG{o}{=} \PYG{n}{gs}\PYG{p}{.}\PYG{n}{getObservation}\PYG{p}{();}
  \PYG{n}{ob}\PYG{p}{.}\PYG{n}{communicationTime}\PYG{p}{;}
  \PYG{n}{ob}\PYG{p}{.}\PYG{n}{computationTime}\PYG{p}{;}
  \PYG{n}{ob}\PYG{p}{.}\PYG{n}{bytesTransmitted}\PYG{p}{;}
\PYG{p}{\PYGZcb{}} \PYG{k}{else} \PYG{p}{\PYGZob{}}
  \PYG{c+c1}{// Sync failed}
\PYG{p}{\PYGZcb{}}
\end{Verbatim}

  \caption{Illustrative usage of \gensync{} from an application.}
  \label{listing:chain}
\end{listing}


\subsection*{Benchmarking Layer}

To allow for performance evaluation under realistic system conditions,
\gensync{} is equipped with a \emph{benchmarking} layer. The
benchmarking layer creates an execution environment,
\mnoter{1.b}\rev{based on the \texttt{cgroups} feature of the Linux
  kernel}, that simulates the target system with a given set of system
parameters. Developers can readily extract the \emph{Observations}
from this simulated environment for further analysis, as we expose
\gensync{}'s benchmarking layer through a script wherein developers
can configure the desired system conditions and the sync protocol to
evaluate (see listing~\ref{listing:benchmarking}).

\begin{listing}[h]

\footnotesize
\begin{Verbatim}[commandchars=\\\{\}]
\PYG{c+c1}{\PYGZsh{} Protocol identifier}
\PYG{n+nv}{protocol}\PYG{o}{=}CPI
\PYG{c+c1}{\PYGZsh{} Latency in milliseconds}
\PYG{n+nv}{latency}\PYG{o}{=}\PYG{l+m}{20}
\PYG{c+c1}{\PYGZsh{} Bandwidth in Mbps (in two directions)}
\PYG{n+nv}{bandwidth}\PYG{o}{=}\PYG{l+s+s2}{\PYGZdq{}10/25\PYGZdq{}}
\PYG{c+c1}{\PYGZsh{} Packet loss (percentage)}
\PYG{n+nv}{packet\PYGZus{}loss}\PYG{o}{=}\PYG{l+m}{0}.01
\PYG{c+c1}{\PYGZsh{} Percentage of CPU cycles used for sync}
\PYG{n+nv}{cpu\PYGZus{}server}\PYG{o}{=}\PYG{l+m}{100}
\PYG{n+nv}{cpu\PYGZus{}client}\PYG{o}{=}\PYG{l+m}{20}
\PYG{c+c1}{\PYGZsh{} Repeat each experiment}
\PYG{n+nv}{repeat}\PYG{o}{=}\PYG{l+m}{100}
\end{Verbatim}

  \caption{\gensync{}'s benchmarking layer configuration script.}
  \label{listing:benchmarking}
\end{listing}




\section*{Included Protocols}\label{sec:included_algos}
\gensync{} includes a number of sync protocols that are based
on compact auxiliary data structures
(\emph{sketches}) through a similar high-level structure. Roughly
speaking, the structure of these protocols consists of
four phases: \circled{1} compute sketches of
local data, \circled{2} exchange the sketches between syncing
hosts, \circled{3} compute
local differences, and \circled{4} exchange differences. The
protocols themselves are distinguished through their choice of sketches and
how the sketches are utilized to infer the differences between the
sets. Specifically, users can currently select from the following protocols.

\algo{CPI} (Characteristic Polynomial Interpolation) sync is based on
a representing data sets as characteristic polynomials. In phase
\circled{1}, both parties encode their elements as zeros of a
characteristic polynomial; they exchange evaluations of these
polynomials in \circled{2}. In \circled{3}, one party extrapolates the
rational function resulting from dividing these polynomials, and
extracts the roots of this function to determine the set differences,
which are then exchanged in \circled{4}.

\algo{Cuckoo} sync uses Cuckoo filters as sketches. In \circled{1},
parties insert all their elements into a Cuckoo filter,
and exchange them in stage \circled{2}. In \circled{3}, each party
queries its own data set elements against the
Cuckoo filter of the other party. Any element for which the Cuckoo
filter returns a negative answer is certainly only locally available
and should thus be sent over in \circled{4}.

\algo{IBLT} (Invertible Bloom Lookup Table) sync uses IBLTs as
sketches, which makes it somewhat similar to Cuckoo. In \circled{1},
each party constructs their own IBLT and exchange them in \circled{2}. In
\circled{3}, one party can subtract the other party's IBLT from their
own to learn the elements that it needs to exchange in \circled{4}.


\begin{figure}[t]
  \centering
  \includegraphics[width=.9\linewidth]{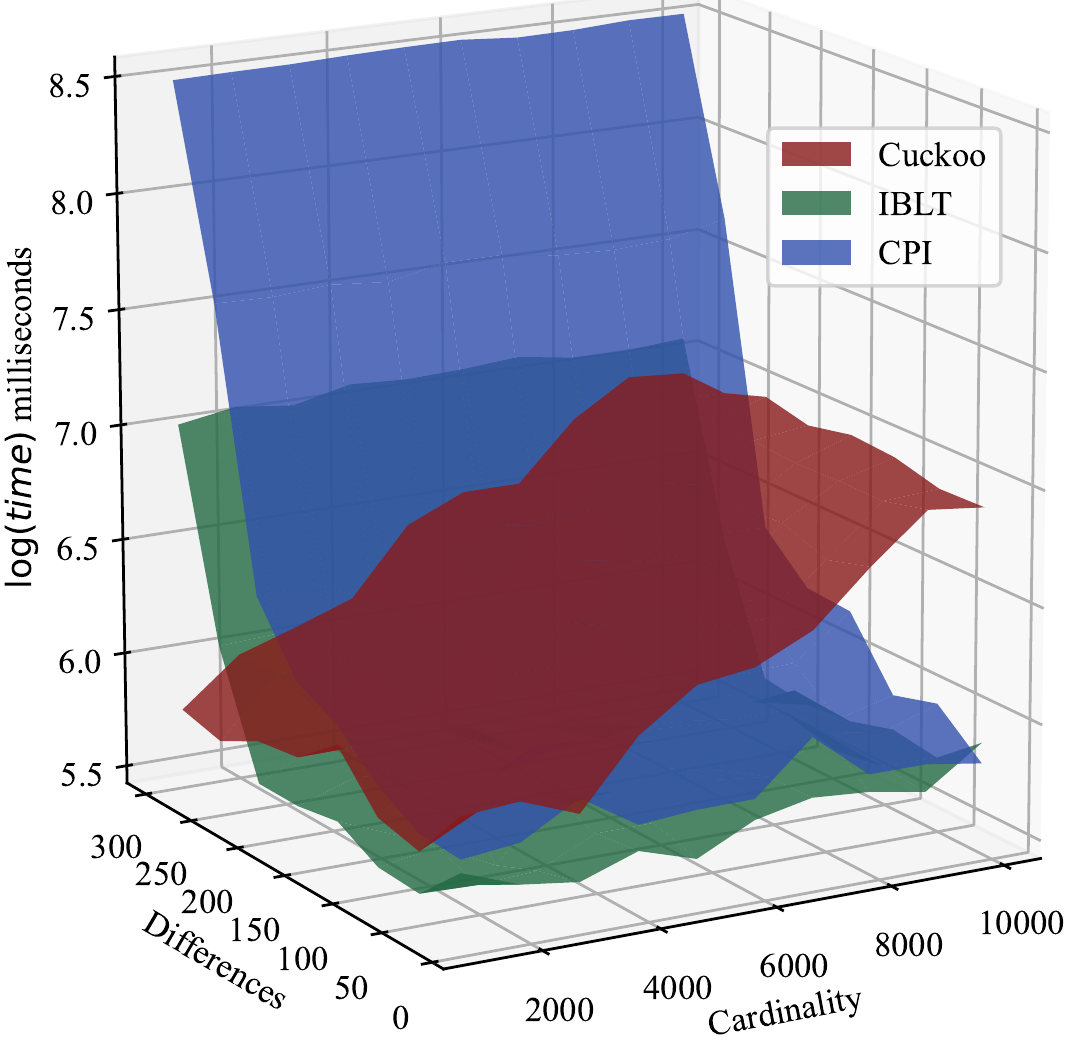}
  \caption{Logarithm of sync time as a function of set cardinality
    (size) and the number of differences.}
  \label{fig:cardinality_diffs_tradeoff}
\end{figure}

\begin{figure}[t]
  \centering
  \includegraphics[width=\linewidth]{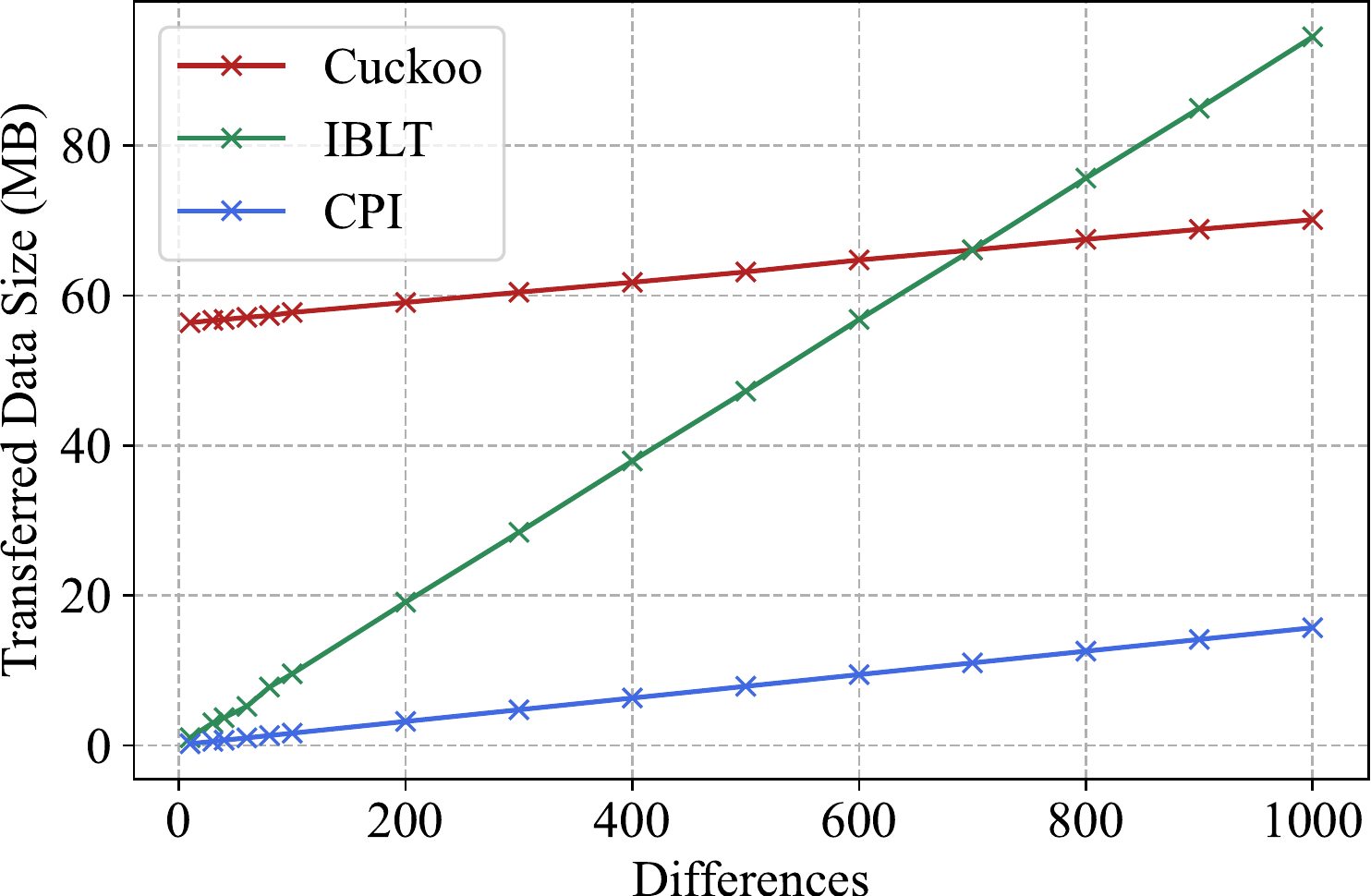}
  \caption{Protocol transfer size as a function of the number of
    mutual differences. Data set size is constant at 10 000.}
  \label{fig:bandwidth_diffs_proj}
\end{figure}

\begin{figure}[t]
  \centering
  \includegraphics[width=\linewidth]{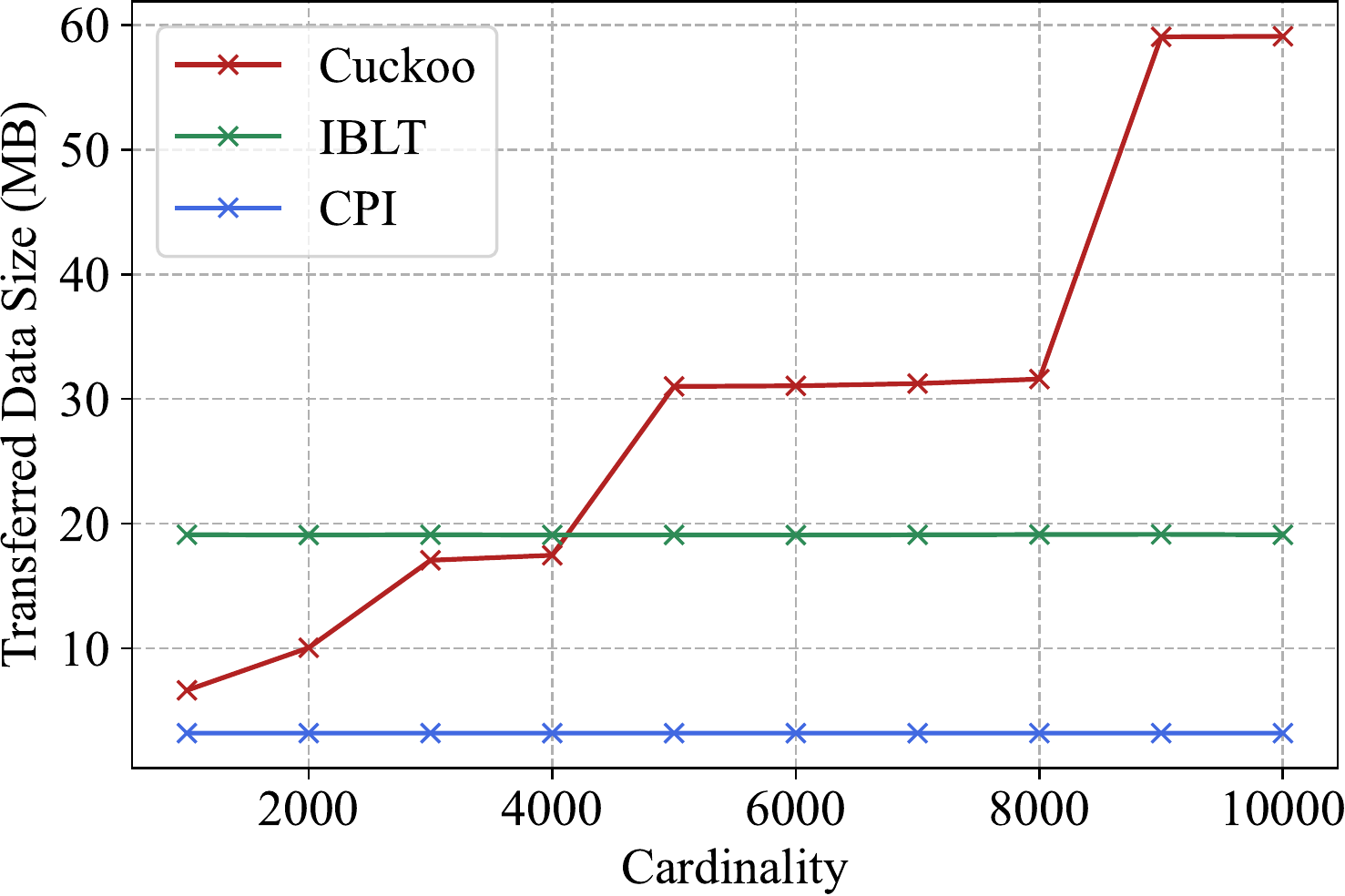}
  \caption{Protocol transfer size as a function of the data set
    size. The number of mutual differences is constant at 100.}
  \label{fig:bandwidth_card_proj}
\end{figure}

\subsection*{Navigating Trade-offs}\label{sec:trade_offs}
Typical performance metrics for data sync protocols are
\emph{transferred data size} (sum of all traffic until the sets are in
sync) and \emph{total sync time}. The transferred data size depends on
\emph{data set parameters} (\ie{} size of the sets and the
number of their mutual differences) and the protocols' theoretical
bounds on communication complexity. The total sync time, however,
depends on \emph{system parameters}, which includes network bandwidth
and latency (jointly referred to as \emph{network conditions}), and
the compute capabilities of the nodes (\emph{compute
  conditions}). Given the systems' complexity, the total sync time
cannot easily be estimated using only the theoretical bounds. Worse
yet, a bad protocol choice for the given system parameters can cause
a 5x loss in the total sync time performance~\cite{gensync}.

Using the \gensync{}'s benchmarking layer, we modeled a
bandwidth-constrained system to explore the effects of data set
parameters on total sync time. We varied set size from ten thousand to
one hundred thousand, and the number of differences between zero and
300. \mnoter{1.b}\rev{We ran \gensync{}'s benchmarking layer on an
  Intel Core i7-7700 experimental server with 5.18.10 version of the
  Linux kernel. The benchmarking layer parameters appear in
  listing~\ref{listing:benchmarking}.} The resulting surfaces are
plotted in \fref{fig:cardinality_diffs_tradeoff}, where we can observe
the following trends.


\takeawaybox{Trend 1}{red!40}{red!10}{The sync time of Cuckoo is
  largely \emph{invariant} to the number of differences.}

In other words, Cuckoo sync performs relatively well for very small sets,
but worsens as the set size increases (regardless of the
differences count). \highlightmnoter{3.g}\highlight{This can be explained with two observations:}
\begin{enumerate*}[label=(\arabic*)]
\item \highlight{we are dealing with a bandwidth-constrained network, and}
\item \highlight{the transfer size for Cuckoo increases (in steps) with the set
  size} (\fref{fig:bandwidth_card_proj}) \highlight{but stays almost constant as
  a function of differences count} (\fref{fig:bandwidth_diffs_proj}).
\end{enumerate*}
The slight increase in transfer size that we observe in
\fref{fig:bandwidth_diffs_proj} is mostly due to the
final transfer of the differences themselves.

\takeawaybox{Trend 2}{red!40}{red!10}{The sync times of IBLT and CPI are
  largely \emph{invariant} to the size of the data sets being synced.}

\begin{figure*}[!t]
  \centering
  \includegraphics[width=\linewidth]{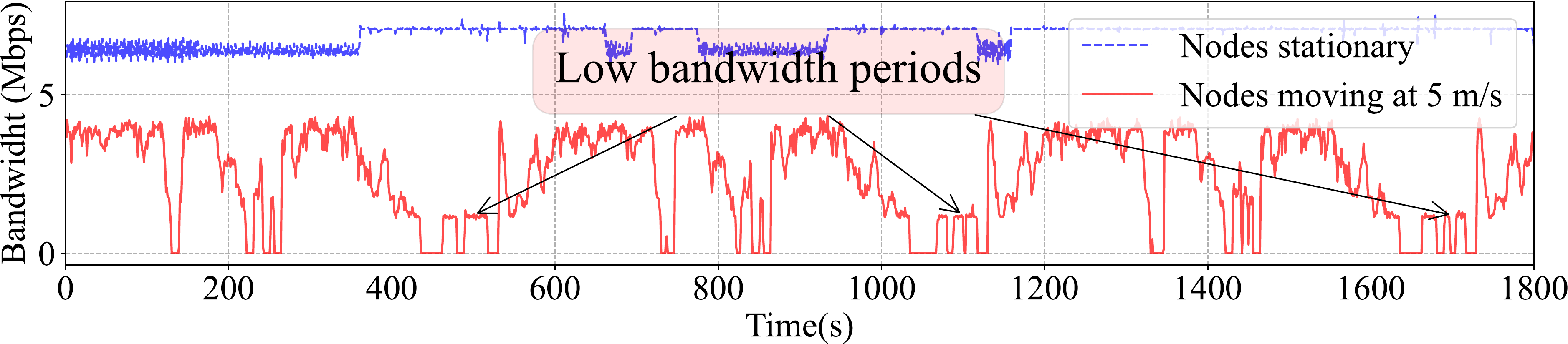}
  \includegraphics[width=\linewidth]{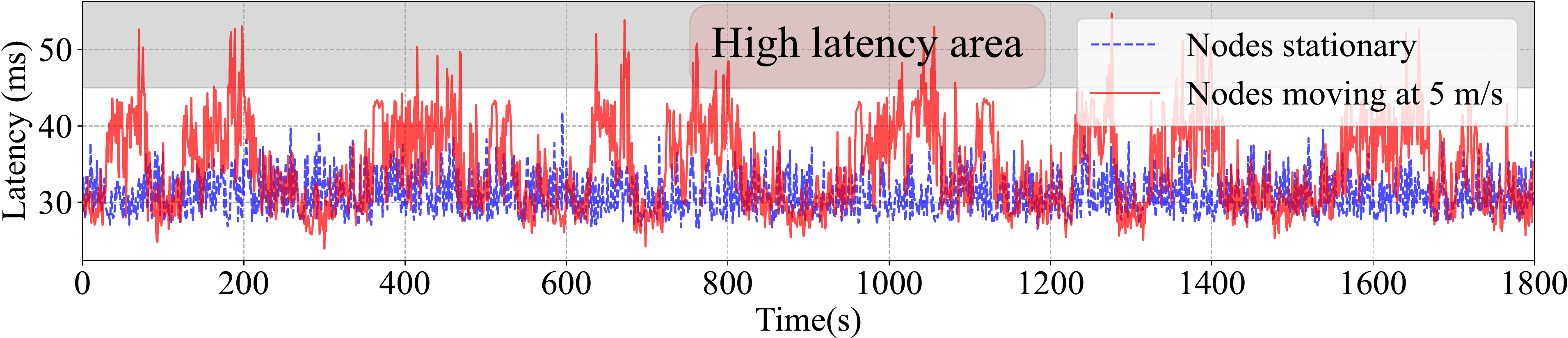}
  \caption{Bandwidth (up) and latency (down) traces from Colosseum~\cite{colosseum},
    one of the world's largest wireless network emulators.}
  \label{fig:colosseum_traces}
\end{figure*}

That is, IBLT and CPI perform well relative to Cuckoo for very similar
sets (\ie{} differing in only a few
elements). \highlightmnoter{3.g}\highlight{IBLT and CPI generally do
  \emph{not} transfer data for the elements that are common.
  Moreover, CPI transfers a nearly optimal amount of data per
  difference~\cite{minsky2003set}, whereas IBLT transfers more}
(\fref{fig:bandwidth_card_proj}). Whether this discrepancy in
transferred data size will result in CPI's dominance (with respect to
total sync time) is the matter of system parameters, and can be
evaluated through \gensync{}'s benchmarking layer.



\section*{Sync on the Edge}\label{sec:case_study}

To showcase the \gensync{} middleware's power in estimating the actual
sync performance in practical systems, we apply it to Colosseum, one
of the world's largest wireless network emulators~\cite{colosseum},
capable of emulating various real world radio frequency
\emph{scenarios} using software-defined radio (SDR) technology. We use
Colosseum's ``Boston'' cellular network scenario to emulate the
cellular network in the vicinity of Boston Common in Boston,
Massachusetts. \rev{The scenario has \emph{stationary} and
  \emph{pedestrian} regimes, where the latter captures moderate user
  movement relative to the base station} (see
Table~\ref{tab:colosseum}).

\begin{table}[h]
  \centering
  \begin{tabular}{>{\centering\arraybackslash}p{1.4cm}>{\centering\arraybackslash}p{1cm}>{\centering\arraybackslash}p{1.6cm}>{\centering\arraybackslash}p{1.4cm}}
    \toprule
    \textbf{Scenario Regime} & \textbf{User} \textbf{Speed}
    & \textbf{Base Station Distance} & \textbf{Scenario Duration}
    \\
    \midrule
    Stationary & \textbf{0 m/s} & 20 m & 600 s \\
    \midrule
    Pedestrian & \textbf{5 m/s} & 20 m & 600 s \\
    \bottomrule
  \end{tabular}
  \caption{\mnote{1.b}\rev{Emulation parameters for the Colosseum's
      Boston scenario~\cite{colosseum}.}}\label{tab:colosseum}
\end{table}

\begin{figure}[t]
  \centering
  \includegraphics[width=\linewidth]{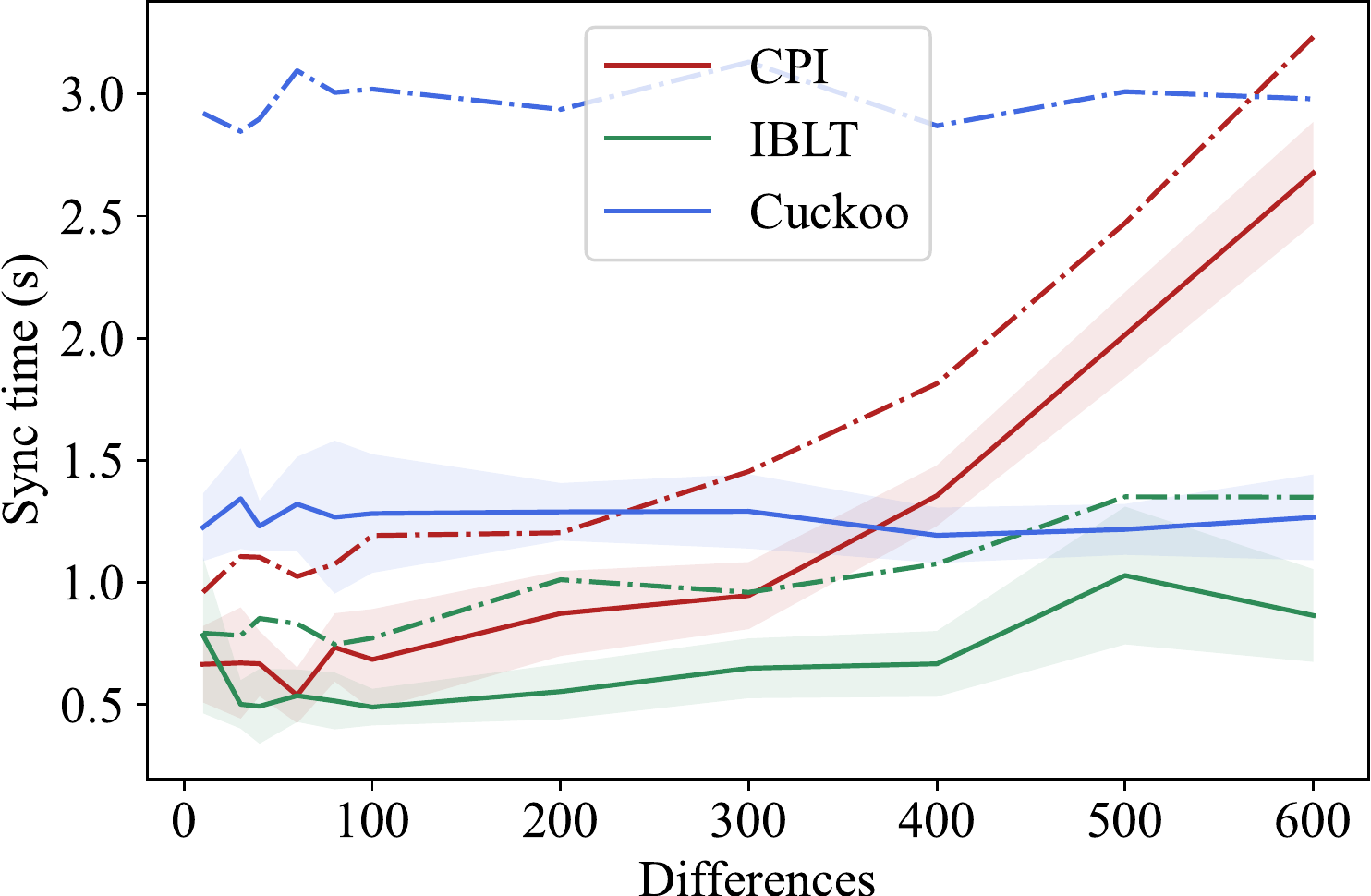}
  \caption{Sync time for the three main sync protocols in \emph{good}
    (solid line) and \emph{bad} (dashed line) network conditions. \rev{Data
      cardinality is $10^8$.} Confidence intervals are
    shaded.}
  \label{fig:colosseum_all}
\end{figure}

The resulting network traces (the available bandwidth and latency) are
plotted in \fref{fig:colosseum_traces}, with their extremes
annotated. These traces show that user movement results in wider oscillations
of available bandwidth and latency. Using the average
values of bandwidth and latency during the extreme periods, we define
two sets of network conditions against which to evaluate our sync protocols:
\begin{enumerate}[label=(\arabic*)]
\item \emph{bad} (bandwidth 1 Mbps, latency 50 ms), and
\item \emph{good} (bandwidth 7 Mbps, latency 30 ms).
\end{enumerate}
The resulting total sync time for the two sets of network conditions
and the three \gensync{} protocols is plotted in
\fref{fig:colosseum_all}. IBLT performs the best in both bad and good
network conditions. \mnoter{3.g}\rev{The reason for this effect lies in
  IBLT's balance of low computational complexity (linear in the number
  of differences) and communication cost that, although higher than
  CPI\cite{minsky2003set}, beats Cuckoo for large data sets. As the
  average bandwidth in both good and bad network conditions does not
  drop below a critical point, IBLT also beats CPI.}

\begin{figure}[h]
  \centering
  \includegraphics[width=\linewidth]{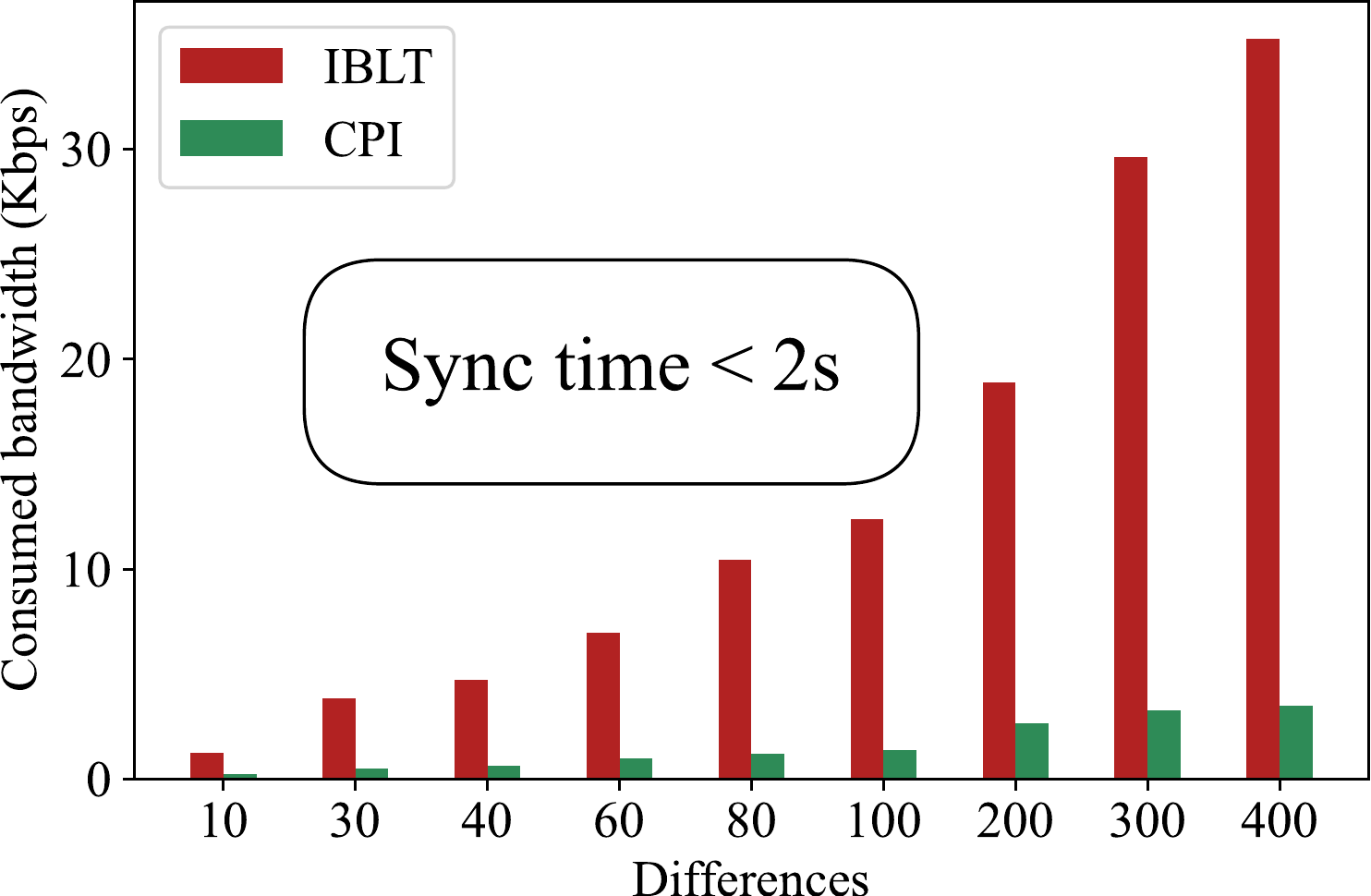}
  \caption{Bandwidth consumed for the two most bandwidth-efficient
    sync protocols that still complete under 2 seconds in \emph{bad}
    network conditions.}
  \label{fig:budgeting}
\end{figure}

However, an application-specific protocol that is being constructed
using \gensync{} may have additional performance objectives. For
instance, it may want to be conservative about the amount of
\emph{consumed bandwidth}, while still keeping a reasonable sync time
(even under bad network conditions). Since the synchronizing device may run
several applications concurrently, this kind of bandwidth budgeting
could be an important dimension to consider when designing an
application-specific sync protocol. Bandwidth savings in the sync
protocol could generate substantial performance gains for Quality-of-Experience-critical
processes, such as streaming point clouds in augmented reality (AR)
applications~\cite{point_clouds}. In \fref{fig:budgeting} we plot the
amounts of bandwidth that IBLT and CPI consume when sync time is
constrained to two seconds.
In this case, CPI achieves almost nine times better performance across
all difference counts. \mnoter{3.g}\rev{The reason for CPI's dominance
  over IBLT in this scenario is CPI's nearly optional communication
  cost, while IBLT adds a multiplicative constant to that cost.}




\section*{Conclusions and Future Goals}\label{sec:conclusion}


The \gensync{} middleware library is the first open and general
framework that\mnoter{1.e}
\begin{enumerate*}[label=(\arabic*)]
\item \rev{enables comparative evaluation of state-of-the-art data
    synchronization protocols in practical environments through a
    versatile benchmarking layer, and}
\item \rev{allows developers to seamlessly integrate the protocol of their
    choice into their applications.}
\end{enumerate*}


\mnote{3.h}\rev{A current limitation of \gensync{} is that is non-adaptive, in the sense
that it cannot intelligently detect the sways in system conditions
(\eg{} available bandwidth) and automatically replace the current
protocol with better one for the new conditions. We leave this
promising direction for future work.}

\rev{We also suggest exploring the extension of \gensync{}'s
protocols to file (bit array) sync, with the hope of improving the
venerable Rsync-based techniques.}


\section*{Acknowledgments}\label{sec:ack}
The authors would like to thank Sevval Simsek for discussions
regarding the Colosseum framework.  The authors would also like to
thank Red Hat, the Boston University Red Hat Collaboratory (award \#
2022-01-RH03), and the US National Science Foundation (award \#
CNS-2210029) for their support.

\bibliographystyle{IEEEtran}
\bibliography{IEEEabrv,bibliography}

\begin{thebibliography}{10}
\providecommand{\url}[1]{#1}
\csname url@samestyle\endcsname
\providecommand{\newblock}{\relax}
\providecommand{\bibinfo}[2]{#2}
\providecommand{\BIBentrySTDinterwordspacing}{\spaceskip=0pt\relax}
\providecommand{\BIBentryALTinterwordstretchfactor}{4}
\providecommand{\BIBentryALTinterwordspacing}{\spaceskip=\fontdimen2\font plus
\BIBentryALTinterwordstretchfactor\fontdimen3\font minus
  \fontdimen4\font\relax}
\providecommand{\BIBforeignlanguage}[2]{{%
\expandafter\ifx\csname l@#1\endcsname\relax
\typeout{** WARNING: IEEEtran.bst: No hyphenation pattern has been}%
\typeout{** loaded for the language `#1'. Using the pattern for}%
\typeout{** the default language instead.}%
\else
\language=\csname l@#1\endcsname
\fi
#2}}
\providecommand{\BIBdecl}{\relax}
\BIBdecl

\bibitem{tridgell}
A.~Tridgell, ``{Efficient Algorithms for Sorting and Synchronization},'' Ph.D.
  dissertation, The Australian National University, 1999.

\bibitem{minsky2003set}
Y.~Minsky, A.~Trachtenberg, and R.~Zippel, ``{Set Reconciliation with Nearly
  Optimal Communication Complexity},'' \emph{IEEE Transactions on Information
  Theory}, vol.~49, no.~9, pp. 2213--2218, 2003.

\bibitem{gensync}
N.~Boškov, A.~Trachtenberg, and D.~Starobinski, ``{GenSync: A New Framework
  for Benchmarking and Optimizing Reconciliation of Data},'' \emph{IEEE
  Transactions on Network and Service Management}, pp. 1--1, 2022.

\bibitem{IoTAccess}
O.~Novo, ``{Blockchain Meets IoT: An Architecture for Scalable Access
  Management in IoT},'' \emph{IEEE Internet of Things Journal}, vol.~5, no.~2,
  pp. 1184--1195, 2018.

\bibitem{blockhainfedsurvey}
W.~Issa, N.~Moustafa, B.~Turnbull, N.~Sohrabi, and Z.~Tari, ``{Blockchain-Based
  Federated Learning for Securing Internet of Things: A Comprehensive
  Survey},'' \emph{ACM Computing Surveys}, August 2022.

\bibitem{block4forensic}
M.~Cebe, E.~Erdin, K.~Akkaya, H.~Aksu, and S.~Uluagac, ``{Block4Forensic: An
  Integrated Lightweight Blockchain Framework for Forensics Applications of
  Connected Vehicles},'' \emph{IEEE Communications Magazine}, vol.~56, no.~10,
  pp. 50--57, 2018.

\bibitem{icnblockchain}
K.~Lei, Q.~Zhang, J.~Lou, B.~Bai, and K.~Xu, ``{Securing ICN-Based UAV Ad Hoc
  Networks with Blockchain},'' \emph{IEEE Communications Magazine}, vol.~57,
  no.~6, pp. 26--32, 2019.

\bibitem{dccs}
S.~Dustdar, V.~Casamajor~Pujol, and P.~K. Donta, ``{On distributed computing
  continuum systems},'' \emph{IEEE Transactions on Knowledge and Data
  Engineering}, pp. 1--1, 2022.

\bibitem{churnAnas2021}
M.~A. {Imtiaz}, D.~{Starobinski}, A.~{Trachtenberg}, and N.~{Younis}, ``{Churn
  in the Bitcoin Network},'' \emph{IEEE Transactions on Network and Service
  Management}, pp. 1--1, 2021.

\bibitem{ozisik2019graphene}
A.~P. Ozisik, G.~Andresen, B.~N. Levine, D.~Tapp, G.~Bissias, and S.~Katkuri,
  ``{Graphene: efficient interactive set reconciliation applied to blockchain
  propagation},'' in \emph{Proceedings of the ACM Special Interest Group on
  Data Communication}, 2019, pp. 303--317.

\bibitem{naumenko2019erlay}
G.~Naumenko, G.~Maxwell, P.~Wuille, A.~Fedorova, and I.~Beschastnikh, ``{Erlay:
  Efficient transaction relay for bitcoin},'' in \emph{Proceedings of the 2019
  ACM SIGSAC Conference on Computer and Communications Security}, 2019, pp.
  817--831.

\bibitem{towardsnetleveff}
Z.~Li, C.~Jin, T.~Xu, C.~Wilson, Y.~Liu, L.~Cheng, Y.~Liu, Y.~Dai, and Z.-L.
  Zhang, ``{Towards Network-Level Efficiency for Cloud Storage Services},'' in
  \emph{Proceedings of the Internet Measurement Conference (IMC)}.\hskip 1em
  plus 0.5em minus 0.4em\relax New York, NY, USA: Association for Computing
  Machinery, 2014, p. 115–128.

\bibitem{wang2018fog}
T.~Wang, J.~Zhou, A.~Liu, M.~Z.~A. Bhuiyan, G.~Wang, and W.~Jia, ``{Fog-based
  computing and storage offloading for data synchronization in IoT},''
  \emph{IEEE Internet of Things Journal}, vol.~6, no.~3, pp. 4272--4282, 2018.

\bibitem{chen2009bdp}
T.~Chen, D.~Guo, X.~Liu, H.~Chen, X.~Luo, and J.~Liu, ``{Bdp: A bloom filters
  based dissemination protocol in wireless sensor networks},'' in \emph{2009
  IEEE 6th International Conference on Mobile Adhoc and Sensor Systems}, 2009,
  pp. 593--602.

\bibitem{Choi2017}
W.~Choi, I.~S. Kim, and D.~H. Lee, ``{E2PKA: An Energy-Efficient and PV-Based
  Key Agreement Scheme for Body Area Networks},'' \emph{Wireless Personal
  Communications}, vol.~97, no.~1, pp. 977--998, Nov 2017.

\bibitem{chronosync}
Z.~Zhu and A.~Afanasyev, ``{Let's ChronoSync: Decentralized dataset state
  synchronization in Named Data Networking},'' in \emph{2013 21st IEEE
  International Conference on Network Protocols (ICNP)}, 2013, pp. 1--10.

\bibitem{chen2014robust}
D.~Chen, C.~Konrad, K.~Yi, W.~Yu, and Q.~Zhang, ``{Robust Set
  Reconciliation},'' in \emph{Proceedings of the ACM International Conference
  on Management of Data (SIGMOD)}, 2014, pp. 135--146.

\bibitem{dccs_rel}
P.~K. Donta and S.~Dustdar, ``{The Promising Role of Representation Learning
  for Distributed Computing Continuum Systems},'' in \emph{2022 IEEE
  International Conference on Service-Oriented System Engineering (SOSE)},
  2022, pp. 126--132.

\bibitem{colosseum}
L.~Bonati, P.~Johari, M.~Polese, S.~D'Oro, S.~Mohanti, M.~Tehrani-Moayyed,
  D.~Villa, S.~Shrivastava, C.~Tassie, K.~Yoder, A.~Bagga, P.~Patel, V.~Petkov,
  M.~Seltser, F.~Restuccia, A.~Gosain, K.~R. Chowdhury, S.~Basagni, and
  T.~Melodia, ``{Colosseum: Large-Scale Wireless Experimentation Through
  Hardware-in-the-Loop Network Emulation},'' in \emph{Proc. of IEEE
  International Symposium on Dynamic Spectrum Access Networks (DySPAN)},
  December 2021.

\bibitem{point_clouds}
L.~Wang, C.~Li, W.~Dai, J.~Zou, and H.~Xiong, ``{QoE-Driven and Tile-Based
  Adaptive Streaming for Point Clouds},'' in \emph{IEEE International
  Conference on Acoustics, Speech and Signal Processing (ICASSP)}, 2021, pp.
  1930--1934.

\end{thebibliography}

\begin{IEEEbiography}{Novak Boškov}{\,}is a Ph.D. candidate at Boston
  University's Department of Electrical and Computer Engineering. His
  research interests include decentralized networks, cloud computing,
  and cybersecurity. Contact him at boskov@bu.edu.
\end{IEEEbiography}


\begin{IEEEbiography}{Ari Trachtenberg}{\,}is a professor at Boston University,
with appointments in Electrical and Computer Engineering, Computer Science,
and Systems Engineering.  His research interests include cybersecurity, distributed
systems, and information theory.  Contact him at trachten@bu.edu.
\end{IEEEbiography}

\begin{IEEEbiography}{David Starobinski}{\,}is a professor at Boston University,
with appointments in Electrical and Computer Engineering, Computer Science,
and Systems Engineering.  His research interests include cybersecurity, wireless networking, and network economics. Contact him at staro@bu.edu.
\end{IEEEbiography}

\end{document}